\documentclass[prc,aps,eqsecnum,amssymb,floatfix,showpacs]{revtex4-1}
\usepackage{ulem} 
\usepackage{dcolumn}
\usepackage{bm}
\usepackage{color}
\usepackage{amssymb}
\usepackage{amsmath}
\usepackage{graphicx}
\usepackage{amsfonts}
\usepackage{slashed}
\usepackage{pstricks}
\usepackage{float} 
\allowdisplaybreaks
\usepackage{morefloats}

\begin{document}
\title{Inclusive neutrino scattering off deuteron at low energies in chiral effective field theory}
\author{A.\ Baroni$^{\,{\rm a}}$ and R.\ Schiavilla$^{\,{\rm a,b}}$}

\affiliation{
$^{\rm a}$\mbox{Department of Physics, Old Dominion University, Norfolk, VA 23529, USA}\\
$^{\rm b}$\mbox{Theory Center, Jefferson Lab, Newport News, VA 23606, USA}\\
}
\date{\today}

\begin{abstract}
Cross sections for inclusive neutrino scattering off deuteron induced by neutral and
charge-changing weak currents are calculated from threshold up to 150 MeV energies
in a chiral effective field theory including high orders in the power counting.  Contributions
beyond leading order (LO) in the weak current are found to be small, and increase
the cross sections obtained with the LO transition operators by a couple of percent
over the whole energy range (0--150) MeV.  The cutoff dependence is negligible, and
the predicted cross sections are within $\sim 2$\% of, albeit consistently larger than,
corresponding predictions obtained in conventional meson-exchange frameworks.
\end{abstract}

\pacs{23.40.Bw,25.30.Pt}

\maketitle
\section{Introduction}
\label{sec:xsect}

A number of studies of neutrino-deuteron scattering were carried out in the past several decades,
and work done up to the mid 1990's is reviewed in Ref.~\cite{Kubodera94}.  In the early 2000's, these
efforts culminated in a set of predictions~\cite{Nakamura01,Nakamura02} for neutrino-deuteron cross
sections induced by both neutral and charge-changing weak currents and incoming neutrino energies
up to 150 MeV.  The calculations were based on the conventional meson-exchange framework, and
used last-generation realistic potentials available at the time and a realistic model for the nuclear weak currents,
which included one- and two-body terms.  The vector part of these currents was shown to provide an excellent
description of the $np$ radiative capture cross section for neutron energies up to 100 MeV~\cite{Nakamura01},
while the axial part was constrained to reproduce the Gamow-Teller matrix element contributing to tritium
$\beta$-decay~\cite{Nakamura02}.  The Nakamura {\it et al.} studies played an important role in the analysis
and interpretation of the Sudbury Neutrino Observatory (SNO) experiments~\cite{Ahmad02}, which have
established solar neutrino oscillations and the validity of the standard model for the generation of energy and
neutrinos in the sun~\cite{Bahcall04}.

Concurrent with those studies was a next-to-next-to-leading order
calculation of neutrino-deuteron cross sections at low energies
($\lesssim 20$ MeV) in an effective field theory in which pion degrees of freedom are integrated out
and which is consequently parametrized in terms of contact terms~\cite{Butler2001}.  In the
strong-interaction sector, the low-energy constants (LECs) multiplying these contact terms were
fixed by fitting the effective range expansions in the $^1$S$_0$ and $^3$S$_1$ two-nucleon channels
(which dominate the low-energy cross sections).  The weak current included one-body
terms with couplings (nucleon magnetic moments and axial coupling constant) taken from experiment 
as well as two-body terms.  In the vector sector, the two LECs associated with these two-body
terms were determined by reproducing the radiative capture rate of neutrons on protons at thermal
energies and the deuteron magnetic moment.  In the axial sector the two-body terms were
characterized by a single LEC (labeled $L_{1,A}$), which however remained undetermined.  Nevertheless,
by fitting the results of Ref.~\cite{Nakamura02}, Butler {\it et al.}~\cite{Butler2001} were able to
show that the resulting value for $L_{1,A}$ was natural, and that the calculated cross
sections reproduced well the energy dependence of those obtained by Nakamura {\it et al.}.

The energy range of the Nakamura {\it et al.}~studies was extended up to 1 GeV in the more
recent calculations by Shen {\it et al.}~\cite{Shen2012}.  These calculations too were based on the
conventional framework, but included refinements in the modeling of the weak currents.  However, they
turned out to have only a minor impact on the predicted cross sections~\cite{Shen2012}.  The results
have confirmed those of Nakamura {\it et al.} in the energy range up to
150 MeV, and have provided important benchmarks for the studies of the weak inclusive response
in light nuclei, including $^{12}$C, with the Green's function Monte Carlo method that have followed
since~\cite{Lovato2013,Lovato2014,Lovato2015,Lovato2016}. They have also been useful in a
recent analysis of the world data on neutrino-deuteron scattering aimed at constraining the
isovector axial form factor of the nucleon~\cite{Meyer2016}, by supplying reliable estimates for
the size of nuclear corrections.

The present study differs from all previous ones in one essential aspect: it is fully based on a
chiral-effective-field-theory ($\chi$EFT) formulation of the nuclear potential~\cite{Entem2003,Machleidt2011}
and weak currents~\cite{Pastore2009,Pastore2011,Piarulli2013,Baroni2016,Baroni2016a}
at high orders in the power counting.  The potential and currents contain intermediate- and
long-range parts mediated by one- and two-pion (and selected multi-pion) exchanges, and a short-range
part parametrized in terms of contact interactions.  The latter are proportional to LECs, which, in the
case of the potential, have been constrained by fitting the nucleon-nucleon scattering database in the
energy range extending up to the pion-production
threshold~\cite{Entem2003,Machleidt2011} and, in the case of the current, by reproducing a number
of low-energy electro-weak observables in the $A\,$=$\, 2$ and 3 nuclei~\cite{Piarulli2013, Baroni2016a}
(specifically, the isoscalar and isovector magnetic moments of the deuteron and trinucleons, and the
tritium Gamow-Teller matrix element).

The importance that accurate predictions for cross sections of neutrino-induced deuteron
breakup into proton-proton and proton-neutron pairs have in the analysis of the SNO experiments,
has prompted us to re-examine these processes in the context of $\chi$EFT.   Because
of its direct connection to the symmetries of quantum chromodynamics, this framework affords
a more fundamental approach to low-energy nuclear dynamics and electro-weak interactions
than the meson-exchange phenomenology adopted in the Nakamura {\it et al.}~\cite{Nakamura01,Nakamura02}
and Shen {\it et al.}~\cite{Shen2012} calculations.  The remainder of this paper is organized as follows.  In
Secs.~\ref{sec:form} and~\ref{sec:cnts} we provide a succinct summary of the theoretical framework,
including the cross section formalism and $\chi$EFT modeling of the nuclear weak currents, while
in Sec.~\ref{sec:res} we present results for the deuteron disintegration cross sections by neutral and
charge-changing weak currents.  A summary and concluding remarks are given in Sec.~\ref{sec:concl}.

\section{Neutrino inclusive cross section}
\label{sec:form}

The differential cross section for neutrino ($\nu$) and antineutrino ($\overline{\nu}$) inclusive scattering off a
deuteron, specifically the processes $^2$H($\nu_l,\nu_l$)$pn$ and $^2$H($\overline{\nu}_l,\overline{\nu}_l$)$pn$
induced by neutral weak currents (NC) and denoted respectively as $\nu_l$-NC and $\overline{\nu}_l$-NC, and the processes $^2$H($\nu_e,e^-$)$pp$ and $^2$H($\overline{\nu}_e,e^+$)$nn$
induced by charge-changing weak currents (CC) and denoted respectively as $\nu_l$-CC and $\overline{\nu}_l$-CC, can be expressed as~\cite{Shen2012}
\begin{equation}
\left(\frac{ {\rm d}\sigma}{ {\rm d}\epsilon^\prime {\rm d}\Omega}\right)_{\nu/\overline{\nu}}=
 \frac{G^2}{8\,\pi^2}\, \frac{k^\prime}{ \epsilon} \,F(Z,k^\prime)\, \Bigg[ v_{00}\, 
R_{00} +v_{zz} \, R_{zz} -v_{0z}\, R_{0z} +
v_{xx}\, R_{xx} \mp v_{xy} \, R_{xy}
\Bigg] \ ,
\label{eq:xswa}
\end{equation}
where $G$=$G_F$ for the NC processes and $G$=$G_F \, {\rm cos}\, \theta_C$ for the
CC processes, and the $-$ ($+$) sign in the last term is relative to the $\nu$ ($\overline{\nu}$)
initiated reactions.  Following Ref.~\cite{Nakamura02}, we adopt the value
$G_F=1.1803\times 10^{-5}$ GeV$^{-2}$ as obtained from an analysis of
super-allowed $0^+ \rightarrow 0^+$ $\beta$-decays~\cite{Towner99}---this
value includes radiative corrections---while ${\rm cos}\, \theta_C$ is taken as 0.97425
from Ref.~\cite{PDG}.  The initial neutrino four-momentum is $k^\mu=(\epsilon, {\bf k})$, the final
lepton four momentum is $k^{\mu \,\prime}=(\epsilon^\prime,{\bf k}^\prime)$, and the
lepton scattering angle is denoted by $\theta$.  We have also defined
the lepton energy and momentum transfers as $\omega=\epsilon-\epsilon^\prime$
and ${\bf q}={\bf k}-{\bf k}^\prime$, respectively, and the squared four-momentum
transfer as $Q^2=q^2-\omega^2 > 0$.  The Fermi function $F(Z,k^\prime)$ with $Z=2$
accounts for the Coulomb distortion of the final lepton wave function in the CC reaction,
\begin{equation}
F(Z,k^\prime) = 2\, (1+\gamma)\, (2\, k^\prime\, r_d)^{2\,\gamma-2}\, {\rm exp} \left(\pi\, y\right)\,
\Bigg| \frac{\Gamma(\gamma+i\, y)}{\Gamma(1+2\,\gamma)} \Bigg|^2 \ , \qquad
\gamma=\sqrt{1-\left(Z\,\alpha\right)^2} \ ,
\end{equation}
and it is set to one otherwise.  Here $y = Z\, \alpha \, \epsilon^\prime/k^\prime$, $\Gamma(z)$ is the
gamma function, $r_d$ is the deuteron charge radius ($r_d=1.97$ fm), and $\alpha$ is the fine
structure constant.

The factors $v_{\alpha\beta}$ denote combinations of lepton kinematical variables including the final
lepton mass, while the nuclear response functions are defined schematically as (explicit expressions
for the $v_{\alpha\beta}$ and $R_{\alpha\beta}$ can be found in Ref.~\cite{Shen2012})
\begin{equation}
\label{eq:r1}
R_{\alpha\beta}(q,\omega) \sim \frac{1}{3} \sum_{M } \sum_f \delta( \omega+m_d-E_f)\,
\langle f| j^\alpha({\bf q},\omega) |d, M \rangle\, \langle f| j^\beta({\bf q},\omega) |d, M \rangle^*  \ ,
\end{equation}
where $|d, M\rangle$ and $|f\rangle$ represent, respectively, the initial deuteron state in spin projection
$M$ and the final two-nucleon state of energy $E_f$, and $m_d$ is the deuteron
rest mass.  The three-momentum
transfer ${\bf q}$ is taken along the $z$-axis (i.e., the spin-quantization axis), and
 $j^\alpha({\bf q},\omega)$ is the time component (for $\alpha=0$) or space
component (for $\alpha=x,y,z$) of the NC or CC, denoted, respectively, by
$j^\alpha_{NC}$ or $j^\alpha_{CC}$.  The former is given by
\begin{equation}
j^\alpha_{NC}=-2\, {\rm sin}^2\theta_W\, j^\alpha_{\gamma, S} + (1-2\, {\rm sin}^2\theta_W) \, j^\alpha_{\gamma, z} 
+\, j^{\alpha 5}_z \ ,
\end{equation}
where $\theta_W$ is the Weinberg angle (${\rm sin}^2\theta_W=0.2312$~\cite{PDG}), $j^\alpha_{\gamma,S}$
and $j^\alpha_{\gamma,z}$ include, respectively,
the isoscalar and isovector terms of the electromagnetic current, and $j^{\alpha 5}_z$ includes the isovector
terms of the axial current (the subscript $z$ on these indicates that they transform as the $z$-component
of an isovector under rotations in isospin space).

The charge-changing weak current is written as the sum of polar- and axial-vector components
\begin{equation}
j^\alpha_{CC}=j^\alpha_{\pm}+j^{\alpha 5}_{\pm} \ , \qquad j_\pm = j_x \pm i\, j_y \ .
\end{equation}
The conserved-vector-current (CVC) constraint relates the polar-vector components $j^\alpha_b$ of the
charge-changing weak current to the isovector component $j^\alpha_{\gamma,z}$ of the electromagnetic
current via
\begin{equation}
\left[ \, T_a \, , \, j^\alpha_{\gamma,z} \, \right]=i\, \epsilon_{azb}\, j^\alpha_b \ ,
\end{equation}
where $T_a$ are isospin operators.  Before turning to a brief discussion of the one- and two-body
$\chi$EFT contributions to the NC and CC, we note that, as described in considerable detail
in Ref.~\cite{Shen2012}, we evaluate, by direct numerical integrations, the matrix elements
of the weak current between the deuteron and the two-nucleon scattering states labeled by the
relative momentum ${\bf p}$ and in given pair-spin and pair-isospin channels, thus avoiding cumbersome
multipole expansions.   Differential cross sections are then obtained by integrating over ${\bf p}$
and summing over the discrete quantum numbers the appropriate matrix-element combinations
entering the response functions~\cite{Shen2012}.

\section{Electro-weak current}
\label{sec:cnts}

\begin{figure}[bth]
\centerline{\includegraphics[width=14cm]{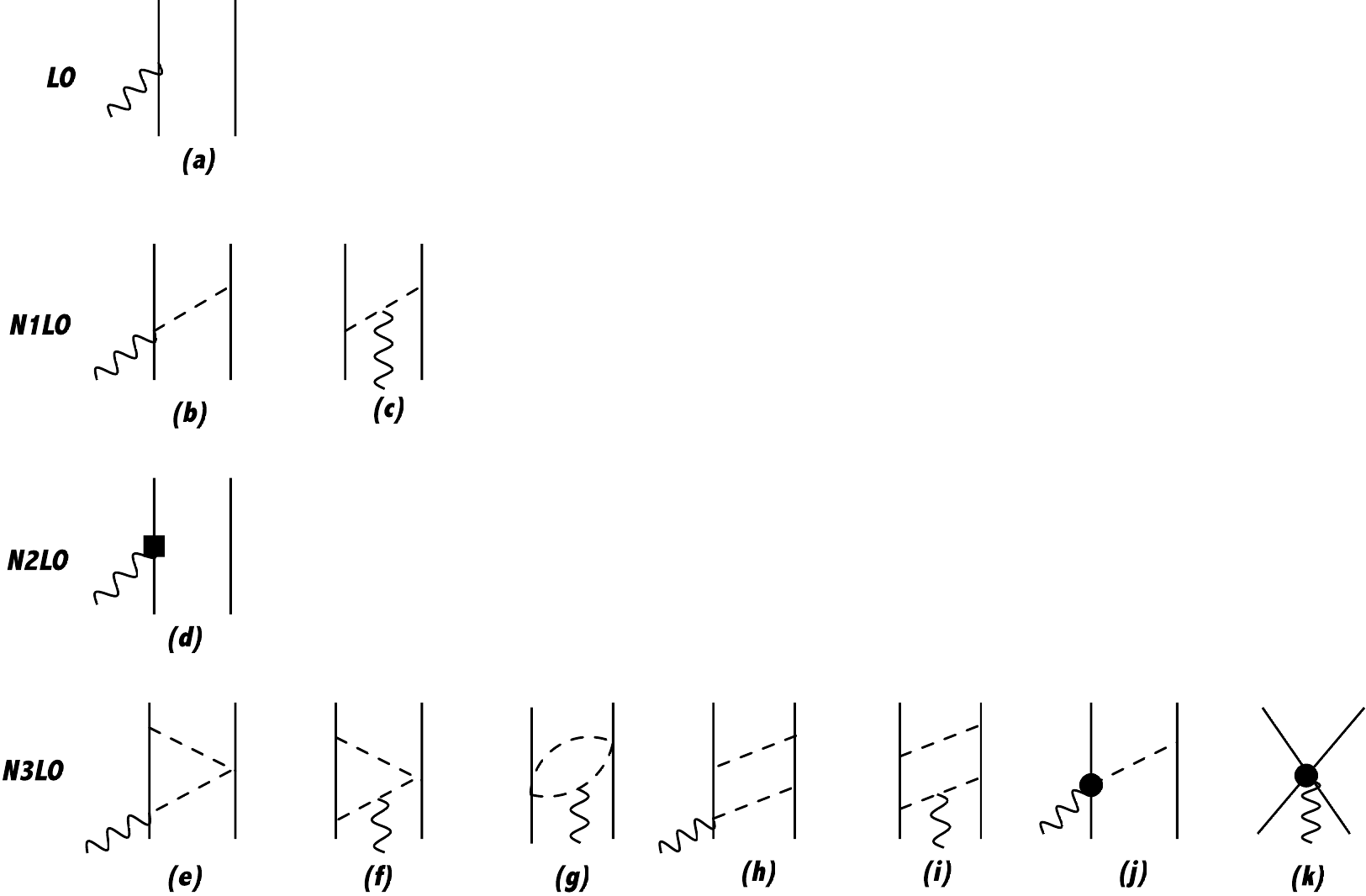}}
\caption{Diagrams illustrating one- and two-body electromagnetic currents entering at $Q^{-2}$ (LO),
$Q^{-1}$ (N1LO), $Q^{\,0}$ (N2LO), and $Q^{\,1}$ (N3LO).  Nucleons, pions, and photons are denoted
by solid, dashed, and wavy lines, respectively.  The square in panel (d) represents the $(Q/m)^2$ relativistic
correction to the LO one-body current ($m$ is the nucleon mass); the solid circle in panel (j) is associated
with the $\gamma \pi N$ coupling involving the LECs $d_8$, $d_9$, and $2\,d_{21}\,$--$\,d_{22}$ in the $\pi N$
chiral Lagrangian ${\cal L}^{(3)}_{\pi N}$~\cite{Fettes2000}; the solid circle in panel (k) denotes two-body
contact terms of minimal and non-minimal nature, the latter involving two unknown LECs (see text).  Only
one among all possible time orderings is shown for the N1LO and N3LO currents, so that all direct- and
crossed-box contributions are accounted for.}
\label{fig:f2}
\end{figure}

\begin{figure}[bth]
\centerline{\includegraphics[width=12cm]{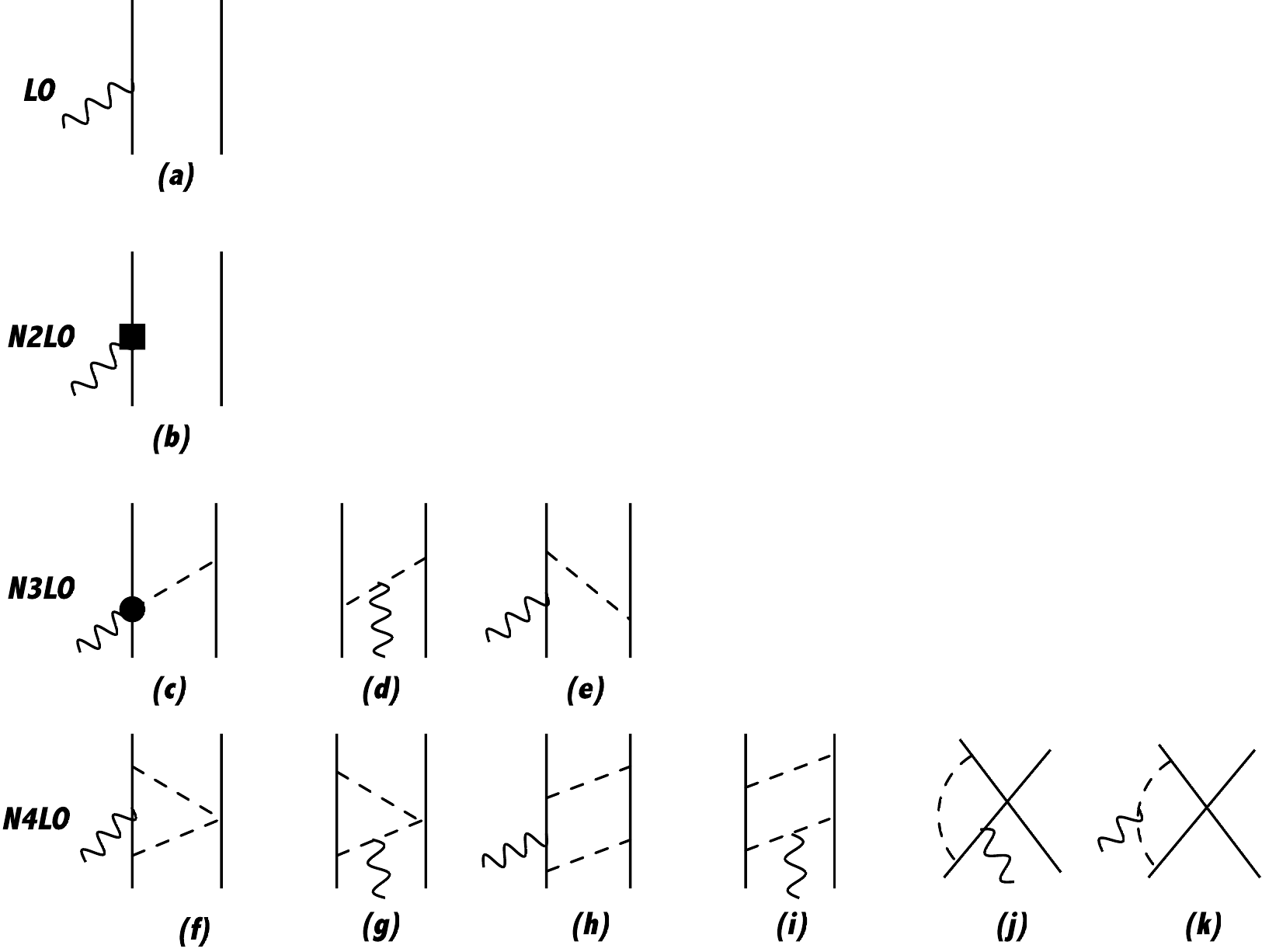}}
\caption{Diagrams illustrating one- and two-body electromagnetic charge operators
entering at $Q^{-3}$ (LO), $Q^{-1}$ (N2LO), $Q^{0}$ (N3LO), $Q^{1}$ (N4LO).
The square in panel (b) represents the $(Q/m)^2$ relativistic correction to the LO
one-body charge operator, whereas panel (c) represents the charge operator $\rho^{(0)}_\gamma({\rm OPE})$
given in Eq.~(\ref{eq:pich}).  As in Fig.~\ref{fig:f2}, only a single time
ordering is shown for the N3LO and N4LO contributions.}
\label{fig:f5}
\end{figure}

\begin{figure}[bth]
\centerline{\includegraphics[width=17cm]{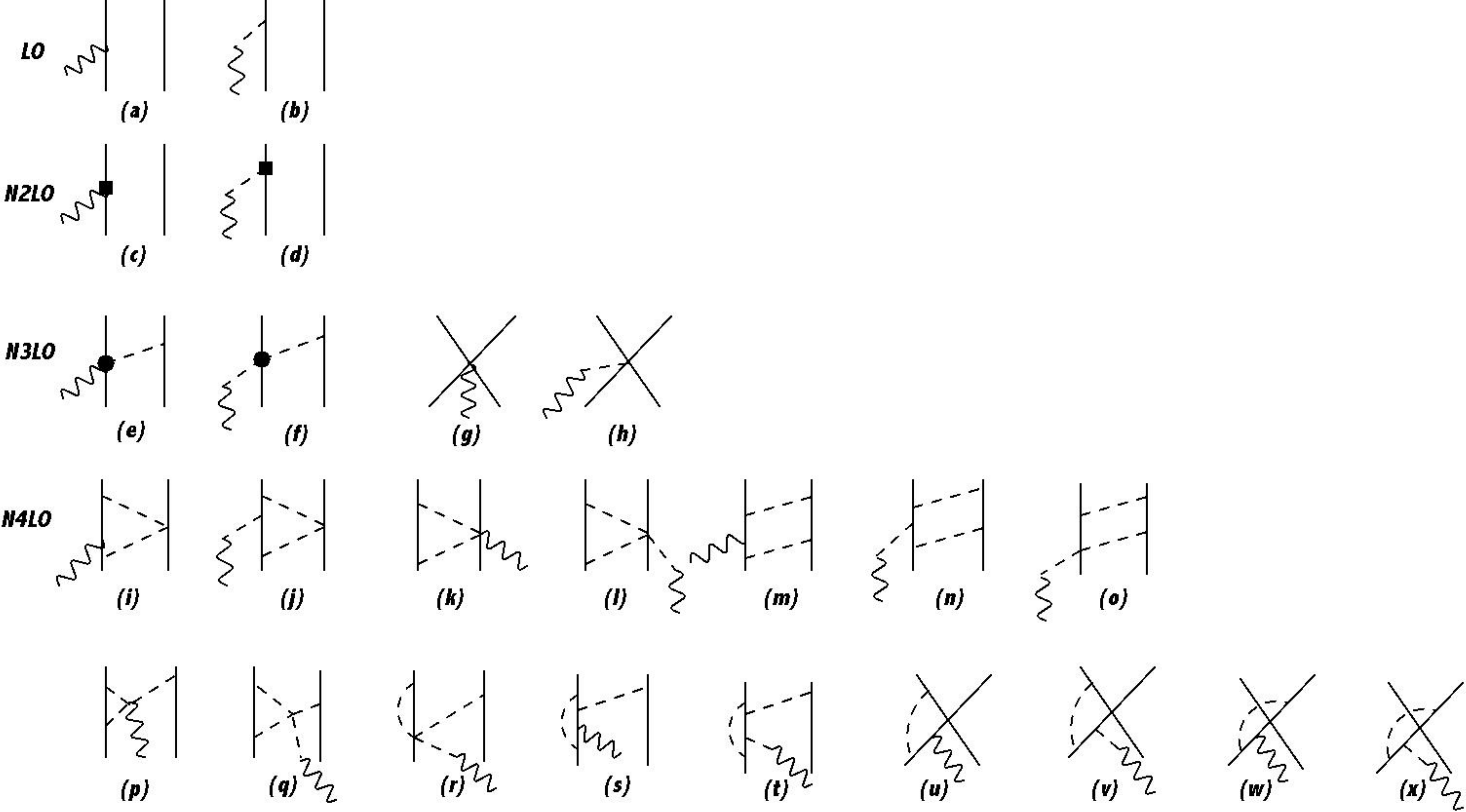}}
\caption{Diagrams illustrating one- and two-body axial currents entering at $Q^{-3}$ (LO),
$Q^{-1}$ (N2LO), $Q^{\,0}$ (N3LO), and $Q^{\,1}$ (N4LO).  Nucleons, pions,
and axial fields are denoted by solid, dashed, and wavy lines, respectively.
The squares in panels (c) and (d) denote relativistic corrections to the one-body
axial current, while the circles in panels (e) and (f) represent vertices implied
by the ${\cal L}^{(2)}_{\pi N}$ chiral Lagrangian~\cite{Fettes2000}, involving
the LECs $c_i$ (see Ref.~\cite{Baroni2016} for additional explanations).
As in Fig.~\ref{fig:f2}, only a single time ordering is shown.}
\label{fig:f2a}
\end{figure}

\begin{figure}[bth]
\centerline{\includegraphics[width=12cm]{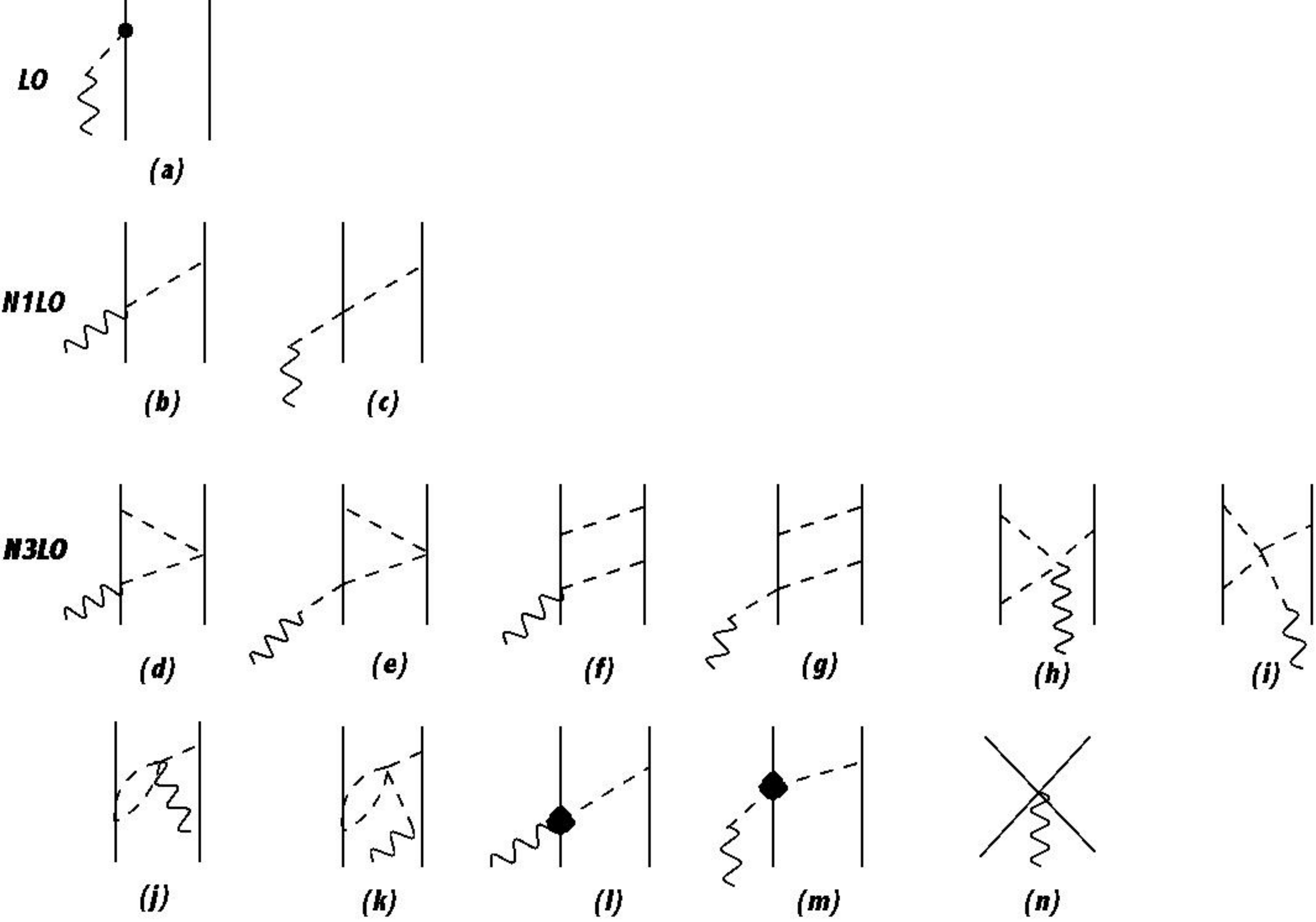}}
\caption{Diagrams illustrating one- and two-body axial charge operators entering
at $Q^{-2}$ (LO), $Q^{-1}$ (N1LO), and $Q^{\,1}$ (N3LO).  Nucleons, pions,
and axial fields are denoted by solid, dashed, and wavy lines, respectively.
The diamonds in panels (l) and (m) indicate higher order $A\pi N$ vertices
implied by the ${\cal L}^{(3)}_{\pi N}$ chiral Lagrangian~\cite{Fettes2000}, involving
the LECs $d_i$ (see Ref.~\cite{Baroni2016} for additional explanations).
As in Fig.~\ref{fig:f2}, only a single time ordering is shown.}
\label{fig:f5a}
\end{figure}

The $\chi$EFT contributions up to one loop to the electromagnetic current~\cite{Pastore2009,Piarulli2013} and
charge~\cite{Pastore2011,Piarulli2013} are illustrated diagrammatically in Figs.~\ref{fig:f2} and~\ref{fig:f5}, while
those to the weak axial current and charge~\cite{Baroni2016,Baroni2016a} in Figs.~\ref{fig:f2a} and~\ref{fig:f5a}.
The former are denoted below as ${\bf j}_\gamma=j^i_\gamma$ and $\rho_\gamma=j^{0}_\gamma$, and the
latter as ${\bf j}_5=j^{i 5}_z$ and $\rho_5=j^{0 5}_z$, respectively, and subscripts specifying isospin components
are dropped for simplicity here.
In these figures, the  N$n$LO corrections are proportional to $Q^{\,n} \times \,Q^{\,\nu_0}$, where
$Q$ denotes generically the low-momentum scale (the expansion parameter is $Q/\Lambda_\chi$,
where $\Lambda_\chi\!\sim\! 1$ GeV is the chiral symmetry breaking scale) and $\nu_0$ characterizes
the leading-order (LO) counting: $\nu_0\,$=$\,-2$ for the electromagnetic current and axial charge and
$\nu_0=-3$ for the electromagnetic charge and axial current [the chiral order in these operators is indicated
by the superscript $(n)$].  We begin by discussing the electromagnetic
operators.

The electromagnetic currents from LO, N1LO, and N2LO terms and from N3LO loop corrections
depend only on the nucleon axial coupling $g_A$ and and pion decay constant $f_\pi$ (N1LO and N3LO),
and the nucleon magnetic moments (LO and N2LO).  Unknown LECs enter the N3LO OPE contribution
involving the $\gamma \pi N$ vertex from the chiral Lagrangian ${\cal L}^{(3)}_{\pi N}$ (see Ref.~\cite{Fettes2000})
as well as the contact currents implied by non-minimal couplings, as discussed in Sec.~\ref{sec:lecs}.  On the other hand, in
the charge operator there are no unknown LECs up to one loop, and OPE contributions, illustrated in
panels (c)-(e) of Fig.~\ref{fig:f5}, only appear at N3LO.
The contributions in panels (d) and (e) involve non-static corrections~\cite{Pastore2011}, while those
in panel (c) lead to the following operator, first derived by Phillips~\cite{Phillips2003},
\begin{equation}
\rho^{(0)}_\gamma({\rm OPE}) =\frac{e\,g_A^2}{8\, m\,f_\pi^2} \left( {\bm \tau}_1 \cdot {\bm \tau_2}
+ \tau_{2z}\right)\, \frac{{\bm \sigma}_1 \cdot {\bf q} \,\, {\bm \sigma}_2 \cdot {\bf k}_2}{k^2_2+m_\pi^2}
+ (1 \rightleftharpoons 2) \ ,
\label{eq:pich}
\end{equation}
where ${\bf q}$ is the momentum imparted by the external field, ${\bf k}_i ={\bf p}_i^\prime -{\bf p}_i$ 
and ${\bf p}_i$ (${\bf p}_i^\prime$) is the initial (final) momentum of nucleon $i$
(with ${\bf k}_1+{\bf k}_2\,$=$\,{\bf q}$), ${\bm \sigma}_i$ and ${\bm \tau}_i$
are its Pauli spin and isospin operators, $m$ ($m_\pi$) is the nucleon (pion) mass.
This operator plays an important role in yielding predictions for the $A\,$=$\,2$--4 charge form factors
that are in excellent agreement with the experimental data at low and moderate values of the momentum
transfer ($q \lesssim 1$ GeV/c)~\cite{Piarulli2013,Marcucci2016}.  The calculations in Ref.~\cite{Piarulli2013} also
showed that the non-static corrections of pion range from panels (d) and (e) of Fig.~\ref{fig:f5}
are typically an order of magnitude smaller than those generated by panel (c).

The axial current and charge operators illustrated in Figs.~\ref{fig:f2a} and~\ref{fig:f5a}
include pion-pole contributions, which are crucial for the current to be conserved
in the chiral limit~\cite{Baroni2016} (obviously, these contributions are suppressed in
low-momentum transfer processes).  There are no direct
couplings of the time-component of the external axial field to the nucleon, see panel (a)
in Fig.~\ref{fig:f5a}.  In the axial current pion-range contributions enter at N3LO, panels
(e) and (f) of Fig.~\ref{fig:f2a}, and involve vertices from the sub-leading ${\cal L}^{(2)}_{\pi N}$
chiral Lagrangian~\cite{Fettes2000}, proportional to the LECs $c_3$, $c_4$, and $c_6$.
The associated operator is given by (the
complete operator, including pion pole contributions, is listed in Ref.~\cite{Baroni2016})
\begin{eqnarray}
\label{eq:opej1fin}
{\bf j}_{5,a}^{(0)}({\rm OPE})&=& \frac{g_A}{2\,f_\pi^2} \bigg\{
2\, c_3 \, \tau_{2,a}\, {\bf k}_2
+\left({\bm \tau}_1\times{\bm \tau}_2\right)_a 
\bigg[ \frac{i}{2\, m} {\bf K}_1 - \frac{c_6+1}{4\, m}  {\bm \sigma}_1\times{\bf q} \nonumber\\
&&+\left( c_4+\frac{1}{4\, m}\right) {\bm \sigma}_1\times{\bf k}_2 \bigg] 
\bigg\} \frac{{\bm\sigma}_2\cdot{\bf k}_2} {k_2^2+m_\pi^2}+ (1\rightleftharpoons 2)\ ,
\end{eqnarray}
where ${\bf K}_i=({\bf p}_i^\prime+{\bf p}_i)/2$.
In contrast, the axial charge has a OPE contribution at N1LO,
illustrated in panels (b) and (c) of Fig.~\ref{fig:f5a}, which reads
\begin{eqnarray}
\label{eq:6.61}
\rho^{(-1)}_{5,a}({\rm OPE})&=&
i\frac{g_A}{4\,f_\pi^2}\left({\bm \tau}_1\times{\bm \tau}_2\right)_a
\frac{{\bm \sigma}_2\cdot{\bf k}_2}{k_2^2+m_\pi^2} + (1\rightleftharpoons 2)\ .
\end{eqnarray}
In fact, an operator of precisely this form was derived by Kubodera {\it et al.}~\cite{Kubodera1978}
in the late seventies, long before the systematic approach based on chiral Lagrangians now in use
had been established.  Corrections to the axial current at N4LO in panels (i)-(x) of Fig.~\ref{fig:f2a}
have been included in a very recent calculation of the tritium Gamow-Teller matrix
element~\cite{Baroni2016a}, while those to the axial charge at N3LO in panels (d)-(n) in
Fig.~\ref{fig:f5a} are considered for the first time in the present study, to the best of our knowledge.
It is worthwhile noting that vertices involving three or four pions, such as those, for example,
occurring in panels (l), (p), (q) and (r) of Fig.~\ref{fig:f2a}, depend on the pion field parametrization.
This dependence must cancel out after summing the individual contributions associated with
these diagrams, as indeed it does~\cite{Baroni2016} (this and the requirement, remarked on below,
that the axial current be conserved in the chiral limit provide useful checks of the calculation).

The loop integrals in the diagrams of Figs.~\ref{fig:f2}--\ref{fig:f5a} are ultraviolet divergent and
are regularized in dimensional regularization~\cite{Pastore2009,Pastore2011,Baroni2016}.
In the electromagnetic current the divergent parts of these loop integrals are reabsorbed by the LECs
multiplying contact terms~\cite{Pastore2009}, while those in the electromagnetic charge cancel out,
in line with the fact that there are no counter-terms at N4LO~\cite{Pastore2011}. In the case of the axial
operators~\cite{Baroni2016}, there are no divergencies in the current, while those in the charge lead to
renormalization of the LECs multiplying contact-type contributions. In particular, the infinities in loop
corrections to the OPE axial charge (not shown in Fig.~\ref{fig:f5a}) are re-absorbed by renormalization
of the LECs $d_i$ in the ${\cal L}^{(3)}_{\pi N}$ chiral Lagrangian.  For a discussion of these issues we
defer to Ref.~\cite{Baroni2016}.

The two-nucleon chiral potentials used in the present study have been derived
up to order $Q^4$~\cite{Entem2003,Machleidt2011}, requiring two-loop contributions.
Conservation of the electromagnetic current ${\bf q}\cdot{\bf j}_\gamma=\left[\, H\, ,\, \rho_\gamma\,\right]$,
where the two-nucleon Hamiltonian is given by $H=T^{(-1)}+v^{(0)}+v^{(2)}+v^{(3)}+v^{(4)}$
with the (two-nucleon) kinetic energy $T^{(-1)}$ being counted as $Q^{-1}$ and where the
$v^{(n)}$'s are the potentials of order $Q^n$, implies~\cite{Pastore2009}, order
by order in the power counting, a set of non-trivial relations between the ${\bf j}_\gamma^{(n)}$ and
the $T^{(-1)}$, $v^{(n)}$, and $\rho_\gamma^{(n)}$.  Since
commutators implicitly bring in extra factors of $Q^{3}$, these relations couple different orders in the power counting
of the operators, making it impossible to carry out a calculation, which at a given $n$ for ${\bf j}_\gamma^{(n)}$,
$v^{(n)}$, and $\rho_\gamma^{(n)}$ (and hence ``consistent'' from a power-counting perspective)
also leads to a conserved current.  Similar considerations also
apply to the conservation of the axial current in the chiral limit~\cite{Baroni2016}.

We conclude this section by noting that a number of independent derivations
of nuclear electromagnetic and axial currents exists in the literature in the $\chi$EFT
formulation adopted here, in which nucleons and pions are the explicit degrees of
freedom. The early and pioneering studies by Park {\it et al.}~\cite{Park1993,Park1996,Park2003}
used heavy-baryon covariant perturbation (HBPT) theory, while the more recent ones by
the Bochum-Bonn group~\cite{Koelling2009,Koelling2011,Krebs2016} are based 
on time-ordered perturbation theory (TOPT) and a different prescription for isolating non-iterative
pieces in reducible diagrams than adopted in Refs.~\cite{Pastore2009,Pastore2011,Piarulli2013,Baroni2016}.
Detailed comparisons between the operators obtained in these latter papers and the HBPT
ones of Park {\it et al.}~can be found in Refs.~\cite{Pastore2009} and~\cite{Baroni2016}.  It suffices
to note here that Park {\it et al.}~in their evaluation of two-nucleon amplitudes have only included
irreducible diagrams and, for the case of the axial currents, did not concern themselves with
pion-pole contributions.  Because of these limitations, the electromagnetic current and axial
current in the chiral limit are not conserved.

The two TOPT-based methods lead to formally equivalent operator structures for the nuclear
potential, electromagnetic current and charge, and axial charge up to one-loop corrections
included~\cite{Piarulli2013}. However, some of the N4LO loop corrections to the axial current
obtained by Krebs {\it et al.}~\cite{Krebs2016} are different from those reported in
Refs.~\cite{Baroni2016,Baroni2016a}.
These differences seem to originate from the evaluation of box diagrams, panels (m) and (n) of Fig.~\ref{fig:f2a}.
Additional differences result from the fact non-static corrections at N4LO have been neglected in
Ref.~\cite{Baroni2016}, while they have been retained explicitly in Ref.~\cite{Krebs2016}.
\subsection{Constraining the LECs in the electro-weak currents}
\label{sec:lecs}

There is a total of ten LECs entering the two-body electro-weak currents discussed above, five
of these are in the electromagnetic (vector) sector and the remaining five (in the limit
of vanishing momentum transfer) in the axial sector.  In
the vector sector, contact terms originate from minimal and non-minimal couplings.  The LECs
multiplying the former are known from fits of the two-nucleon scattering database~\cite{Piarulli2013}.
Non minimal couplings enter through the electromagnetic field tensor, and it has been shown~\cite{Pastore2009}
that only two independent structures occur at order $Q^1$ (see panel (k) in Fig.~\ref{fig:f2}):
\begin{equation}
{\bf j}^{(1)}_{\gamma}({\rm CT})= - i\,e \Big[ \widetilde{c}_\gamma^{\, S}\, {\bm \sigma}_1 
+\widetilde{c}_\gamma^{\, V} (\tau_{1,z} - \tau_{2,z})\,{\bm \sigma}_1  \Big]\times {\bf q}  
+ \left(1 \rightleftharpoons 2\right)\ ,
\label{eq:gnm}
\end{equation}
where $e$ is the electric charge, $\widetilde{c}_\gamma^{\,S}$ and $\widetilde{c}_\gamma^{\,V}$ are the two LECs, and
the superscripts specify the isoscalar ($S$) and isovector ($V$) character
of the associated operator. There is also a pion-range two-body operator
resulting from sub-leading $\gamma \pi N$ couplings associated with the ${\cal L}^{(3)}_{\pi N}$
Lagrangian, and illustrated by panel (j) in Fig.~\ref{fig:f2}.  It reads:
\begin{eqnarray}
\label{eq:cdlt}
{\bf j}_\gamma^{(1)}({\rm OPE})&=& i\,e\, \frac{g_A}{4\,f_\pi^2} \,
\frac{{\bm \sigma}_2 \cdot {\bf k}_2}{k_2^2+m_\pi^2} \bigg[ \left(\widetilde{d}^{\, V}_{\gamma,1} \tau_{2,z}
+ \widetilde{d}_\gamma^{\, S} \, {\bm \tau}_1\cdot{\bm \tau}_2 \right){\bf k}_2\nonumber \\
&&-\widetilde{d}^{\, V}_{\gamma,2} ({\bm \tau}_1\times{\bm \tau}_2)_z\, {\bm \sigma}_1\times {\bf k}_2  \bigg] \times {\bf q} + (1\rightleftharpoons 2) \ ,
\end{eqnarray}
where the LECs $\widetilde{d}_{\gamma,1}^{\, V}$, $\widetilde{d}_{\gamma,2}^{\, V}$ and $\widetilde{d}_\gamma^{\, S}$
are related~\cite{Piarulli2013} to the LECs $d_8$, $d_9$, $d_{21}$, and $d_{22}$ in the
original ${\cal L}^{(3)}_{\pi N}$ Lagrangian~\cite{Fettes2000} in the following way
\begin{equation}
\widetilde{d}_{\gamma}^{\, S}=-8\,d_9\ ,\qquad
\widetilde{d}_{\gamma,1}^{\, V}=-8\,d_8\ ,\qquad
\widetilde{d}_{\gamma,2}^{\, V}=2\,d_{21}-d_{22}\ .
\end{equation}
 As discussed below, these LECs
have been determined by a combination of resonance saturation arguments and fits to photo-nuclear
data in the two- and three-nucleon systems.

In the weak axial sector, there is a single contact term at order $Q^0$ (or N3LO, see panels
(g) and (h) of Fig.~\ref{fig:f2a})
\begin{equation}
\label{eq:jctct}
{\bf j}_{5,a}^{(0)}({\rm CT})=\widetilde{c}^{\, V}_{5,1}\left({\bm \tau}_1\times{\bm \tau}_2\right)_a \left[{\bm \sigma}_1\times{\bm \sigma}_2-\frac{{\bf q}}{q^2+m_\pi^2}\, {\bf q}\cdot\left({\bm \sigma}_1\times{\bm \sigma}_2\right) \right]\ ,
\end{equation}
where the second term of Eq.~(\ref{eq:jctct}) is the pion-pole contribution,
and none at order $Q^1$ (or N4LO).
The axial charge operators at N3LO from OPE [panels (l) and (m) of Fig.~\ref{fig:f5a}] and contact
interactions [panel (n)] involve, in principle, nine LECs~\cite{Baroni2016}.
Since the processes of interest in the present work are relatively low-momentum transfer ones, however,
we have considered here these operators in the limit $q \rightarrow 0$ (or ${\bf k}_1 \simeq -{\bf k}_2$),
which leads to
\begin{eqnarray}
\label{eq:6.66}
\!\!\!\!\!\!\! \rho^{(1)}_{5,a}({\mbox{OPE}})\!&=&\!
i\frac{g_A}{384\,\pi^2\,f_\pi^{4}}\left({\bm \tau}_1\times{\bm \tau}_2\right)_a
\bigg\{g_A^2 \left [\left(5\, k_2^2+8\, m_\pi^{2}\right)\frac{s_2}{k_2}
\ln\frac{s_2+k_2}{s_2-k_2}-\frac{13}{3}k_2^2+2\,m_\pi^2\right] \nonumber\\
\!\!\!\!\!\!\!\!&+& \!\!\left(\frac{s_2^3}{k_2}\ln\frac{s_2+k_2}{s_2-k_2}-\frac{5}{3}k_2^2-8\, m_\pi^{2} \right)
+\widetilde{d}^{\, V}_{5,1}\, k_2^2
+ \widetilde{d}^{\, V}_{5,2} \,m_\pi^{2}\bigg\}\frac{{\bm \sigma}_2\cdot{\bf k}_2}{k_2^2+m_\pi^2} + (1\rightleftharpoons 2)\ , 
\end{eqnarray}
\begin{eqnarray}
\rho^{(1)}_{5,a}({\rm CT}) &=& i\, \widetilde{c}^{\, V}_{5,2} \left({\bm\tau}_1\times{\bm\tau}_2\right)_a 
\, {\bm \sigma}_1\cdot{\bf k}_1
+ i\, \widetilde{c}^{\, V}_{5,3}\, \tau_{1,a}\, \left({\bm \sigma}_1\times{\bm \sigma}_2\right)\cdot {\bf k}_2+ (1\rightleftharpoons 2)\ ,
\label{eq:4.22}
\end{eqnarray}
where $s_j=\sqrt{k_j^2+4\, m_\pi^2}$.  The LECs $\widetilde{d}^{\, V}_{5,i}$ denote the
combinations~\cite{Baroni2016}
\begin{equation}
\widetilde{d}^{\, V}_{5,1}= 4 \,(d_1+d_2+d_3) \ , \qquad \widetilde{d}^{\, V}_{5,2} = 4\, ( d_1+d_2+d_3)+8 \, d_5 \ ,
\end{equation}
in terms of the $d_i$'s in ${\cal L}^{(3)}_{\pi N}$~\cite{Fettes2000}, and
are taken from an analysis of $\pi N$ scattering data as reported in Ref.~\cite{Machleidt2011}.
(It should be noted that a new analysis of these data has
become recently available~\cite{Hoferichter2015}.)  The LECs
$\widetilde{c}^{\, V}_{5,2}$ and $\widetilde{c}^{\, V}_{5,3}$
have yet to be determined.

Configuration-space representations of the $\chi$EFT operators in Figs.~\ref{fig:f2}--\ref{fig:f5a} are required
in the computer programs.  Those for the one-body operators, illustrated in panels
(a) and (d) in Fig.~\ref{fig:f2}, (a) and (b) of Fig.~\ref{fig:f5}, (a)-(d) of Fig.~\ref{fig:f2a},
and (a) of Fig.~\ref{fig:f5a}, follow directly from the momentum-space expressions
listed in Refs.~\cite{Piarulli2013,Baroni2016} by simply multiplying each term in these expressions
by ${\rm exp}(i{\bf q}\cdot{\bf r}_i)$ and by replacing ${\bf K}_i$ with $ -i\, {\bm \nabla}_i$
(and properly symmetrizing for hermiticity).  The configuration-space representations
of the two-body operators are strongly singular at short inter-nucleon separations and must
be regularized before they can be sandwiched between nuclear wave functions.
This is accomplished by insertion in the Fourier transforms of a regulator of the form
$C_\Lambda (k)={\rm exp}[-(k/ \Lambda)^n]$ with $n\,$=$\,4$
and $\Lambda$ in the range (500--600) MeV.  For processes involving low momentum
and energy transfers one would expect predictions to be fairly insensitive to variations
of $\Lambda$, and this expectation is indeed borne out by the calculations reported in the
present work.

\begin{table}[bth]
\caption{The LECs in units of powers of $1/\Lambda$ ($\Lambda$ is the short-range
cutoff) as in Eq.~(\ref{eq:adi}).  Their values are adimensional.  See text for forther explanations. }
\begin{tabular}{c|ccc|cc||cc|ccc}
\toprule
$\Lambda$ (MeV)  & $d_\gamma^S$  & $d^V_{\gamma,1}$ & $d^V_{\gamma,2}$ & $c_\gamma^S$ & $c_\gamma^V$ & $d^V_{5,1}$ & $d^V_{5,2}$ &$c_{5,1}^V$ & $c_{5,2}^V$ &  $c_{5,3}^V$ \\
\hline
500  &  0.219  & 3.458 & 0.865  & 4.072   & --7.981   & --0.210  & 0.690  & 13.22  &0.062 & 0.062 \\
600  &  0.323 & 4.980 & 1.245  & 11.38  & --11.69  & --0.302  &  0.994 &25.07 & 0.130 & 0.130 \\
\botrule
\end{tabular}
\label{tb:t1}
\end{table}
In the electromagnetic sector, the two isoscalar LECs $\widetilde{c}_\gamma^{\, S}$ and
$\widetilde{d}_\gamma^{\, S}$ are fixed (for each $\Lambda$) by reproducing the deuteron and isoscalar
trinucleon magnetic moments, while the two isovector LECs $\widetilde{d}^{\, V}_{\gamma,1}$
and $\widetilde{d}^{\, V}_{\gamma,2}$ are constrained by assuming $\Delta$-resonance
saturation~\cite{Piarulli2013},
\begin{equation}
\widetilde{d}^{\, V}_{\gamma,1}=\frac{4\, \mu_{\gamma N\Delta} \, h_A }{9\, m\,( m_\Delta-m)} \ , \qquad
\widetilde{d}^{\, V}_{\gamma,2}=\frac{1}{4}\, \widetilde{d}^{\, V}_{\gamma,1}\ ,
\end{equation}
where $m_\Delta\,$--$\,m\,$=$\, 294$ MeV, $h_A/(2 f_\pi)=f_{\pi N\Delta}/m_\pi$
with $f^2_{\pi N\Delta}/(4\,\pi)=0.35$ as obtained by equating the first-order
expression of the $\Delta$-decay width to the experimental value, and the transition magnetic
moment $\mu_{\gamma N\Delta}$ is taken as $3\, \mu_N$~\cite{Carlson1986}.  The remaining
LEC $\widetilde{c}_\gamma^{\, V}$ is determined by reproducing
the isovector trinucleon magnetic moment~\cite{Piarulli2013}.  In the weak axial sector, the LEC
$\widetilde{c}^{\, V}_{5,1}$ is fixed by reproducing the tritium Gamow-Teller
matrix element~\cite{Baroni2016a}, while the other two LECs $\widetilde{c}^{\, V}_{5,2}$
and $\widetilde{c}^{\, V}_{5,3}$ in the axial charge are taken here to assume natural values
$\widetilde{c}^{\, V}_{5,i} \simeq 1/\Lambda_\chi^4$,
for $i\,$=$\,2,3$ and with $\Lambda_\chi\,$=$\,1$ GeV.  However,
cross sections results are insensitive to variations of $\widetilde{c}^{\, V}_{5,2}$ and
$\widetilde{c}^{\, V}_{5,3}$ over a rather broad range (see Sec.~\ref{sec:res}).  In Table~\ref{tb:t1} we
list the values of all these LECs in units of the short-range cutoff $\Lambda$, namely
\begin{eqnarray}
\label{eq:adi}
&&\widetilde{d}_\gamma^{\,S}=d^{\, S}_\gamma/\Lambda^2\ ,\qquad \widetilde{d}_{\gamma,i}^{\,V}=d^{\, V}_{\gamma,i} /\Lambda^2\ , 
\qquad \widetilde{c}_\gamma^{\, S}=c^{\, S}_\gamma/\Lambda^4\ ,\qquad \widetilde{c}_\gamma^{\, V}=c^{\,V}_\gamma/\Lambda^4 \ ,\nonumber \\
&& \widetilde{d}_{5,i}^{\, V}=d^{\,V}_{5,i}/\Lambda^2 \ ,\qquad
\widetilde{c}_{5,1}^{\, V}=c^{\,V}_{5,1}/\Lambda^3 \ ,\qquad
  \widetilde{c}_{5,2}^{\, V}=c^{\,V}_{5,2}/\Lambda^4 \ ,
\qquad  \widetilde{c}_{5,3}^{\, V}=c^{\,V}_{5,3}/\Lambda^4 \ .
\end{eqnarray}

Finally, we note that, since the processes under consideration involve small but
non-vanishing four-momentum transfers $Q^2$, hadronic electro-weak form factors
need to be included in the $\chi$EFT operators.  Some of these form factors have
been calculated in chiral perturbation theory~\cite{Kubis2001}, but the
convergence of this calculation in powers of the momentum transfer appears
to be rather poor.  For this reason, in the results reported below, the form
factors in the electromagnetic current and charge are accounted for as in Ref.~\cite{Piarulli2013},
{\it i.e.}, the nucleon, pion, and $N\Delta$-transition electromagnetic form factors
are taken from fits to available electron scattering data.
For the case of the axial charge and current, the operators are simply multiplied by
$G_A(Q^2)/g_A$, where $G_A(Q^2)$ is the nucleon axial form factor, parametrized as
$G_A(Q^2)=g_A/(1+Q^2/\Lambda^2_A)^2$ with $\Lambda_A\,$=$\, 1$ GeV, consistently
with available neutrino scattering data (see~\cite{Shen2012} and references therein).
 
\section{Cross-section predictions}
\label{sec:res}
\begin{figure}[bth]
\centerline{\includegraphics[width=16cm]{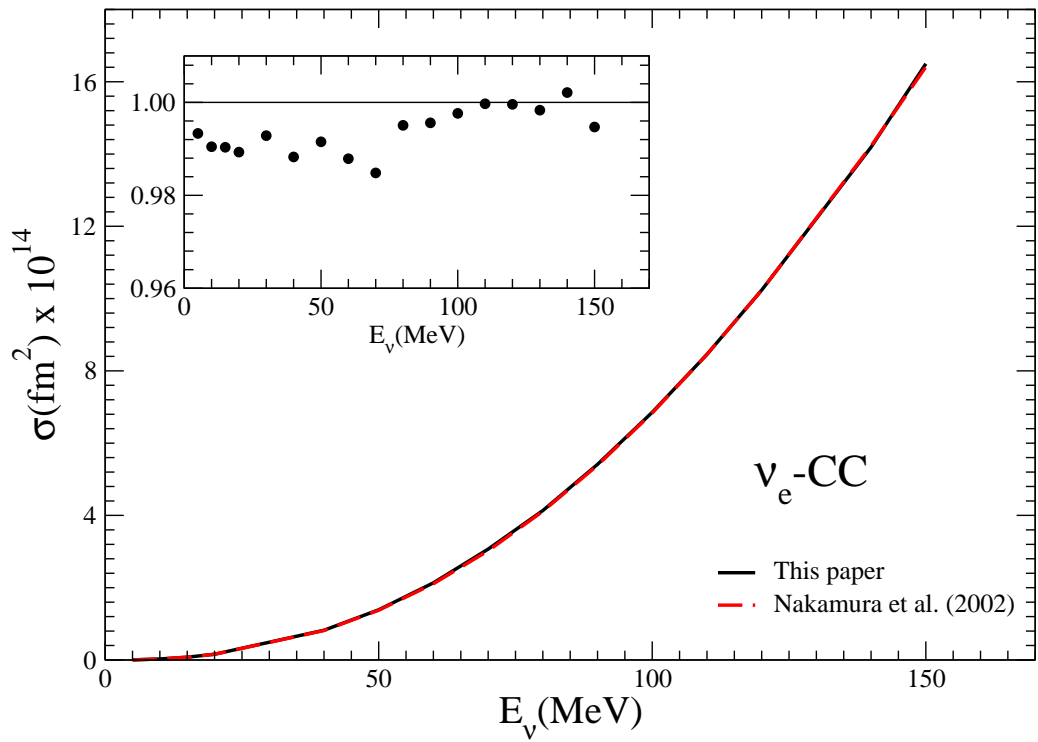}}
\caption{(Color online) Total cross sections in ${\rm fm}^2$ for the $\nu_e$-CC induced process on the deuteron.
The solid line corresponds to the $\chi$EFT calculation with cutoff $\Lambda=$ 500 MeV, based on the chiral
potential of Ref.~\cite{Entem2003} and including electro-weak contributions up to N3LO in the vector current
and axial charge, and up to N4LO in the axial current and vector charge, see Figs.~\ref{fig:f2}--\ref{fig:f5a}.
The dashed line is obtained within the conventional meson-exchange picture of Ref.~\cite{Nakamura02}.
The inset shows the ratio of conventional to $\chi$EFT predictions.}
\label{fig:nu-pp}
\end{figure}

\begin{figure}[bth]
\centerline{\includegraphics[width=16cm]{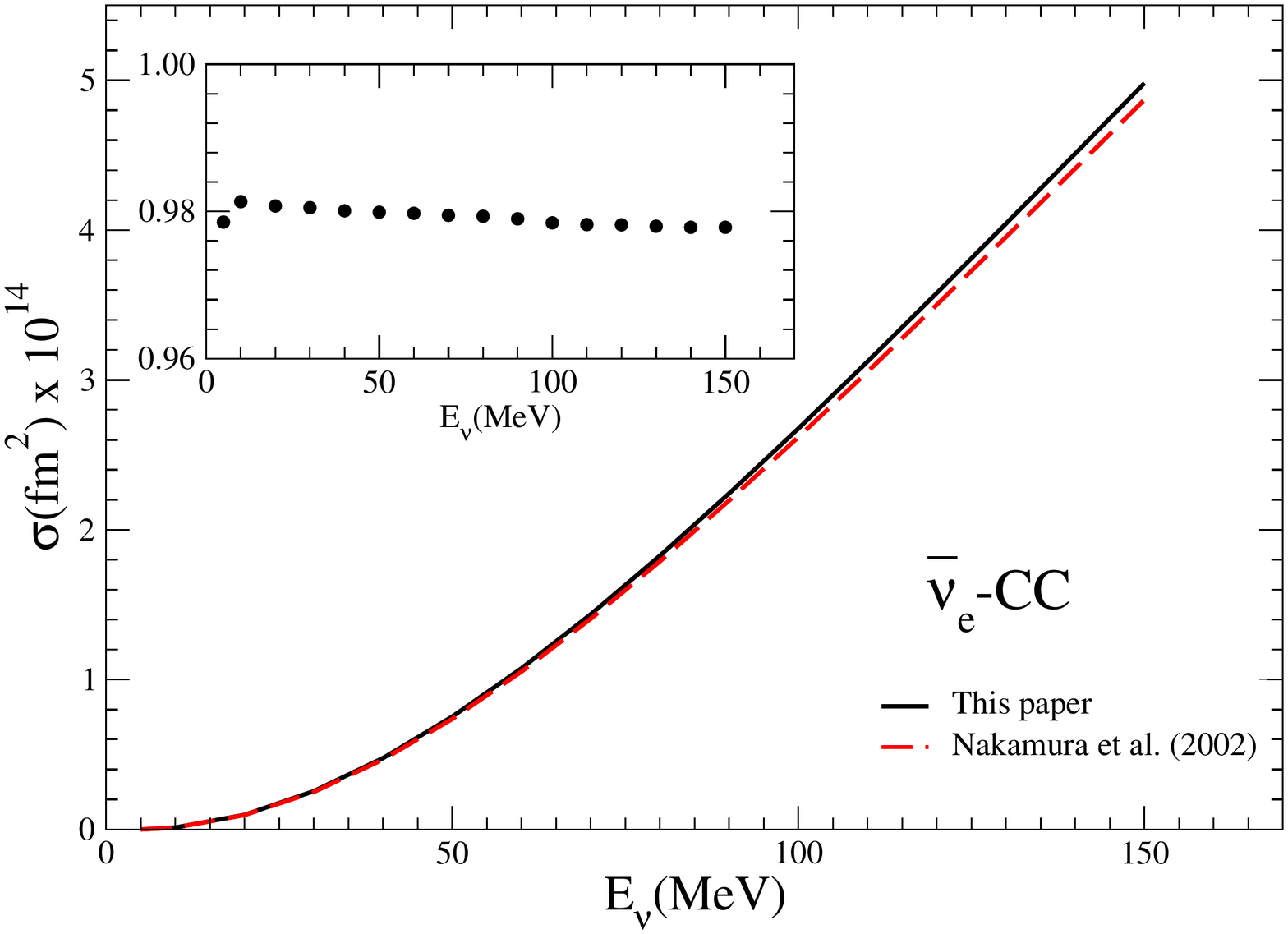}}
\caption{(Color online) Same as in Fig.~\ref{fig:nu-pp} but for the $\overline{\nu}_e$-CC induced process on the deuteron.}
\label{fig:nu-nn}
\end{figure}

\begin{figure}[bth]
\centerline{\includegraphics[width=16cm]{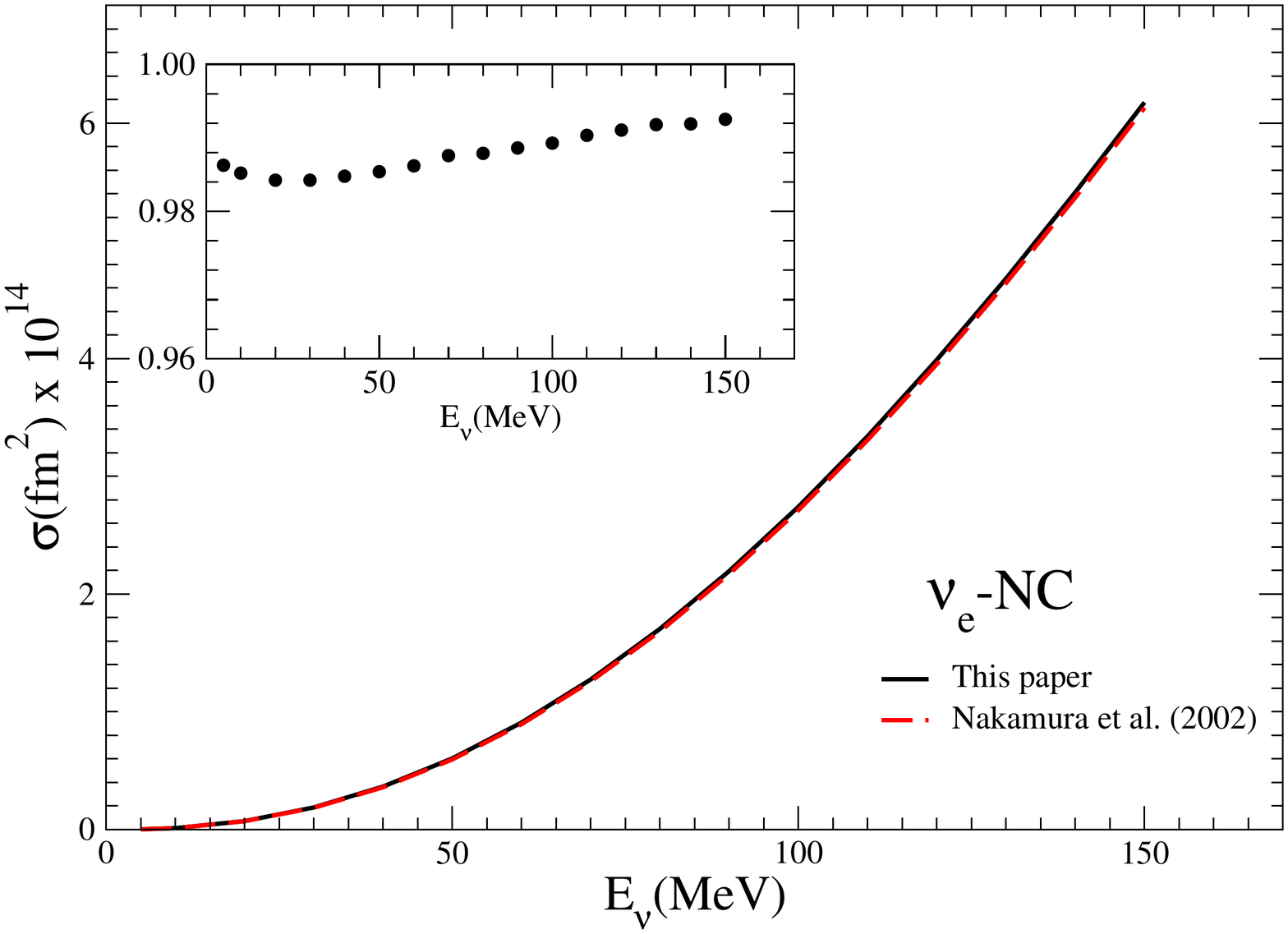}}
\caption{Same as in Fig.~\ref{fig:nu-pp} but for the $\nu_e$-NC induced process on the deuteron.}
\label{fig:nu-np}
\end{figure}

\begin{figure}[bth]
\centerline{\includegraphics[width=16cm]{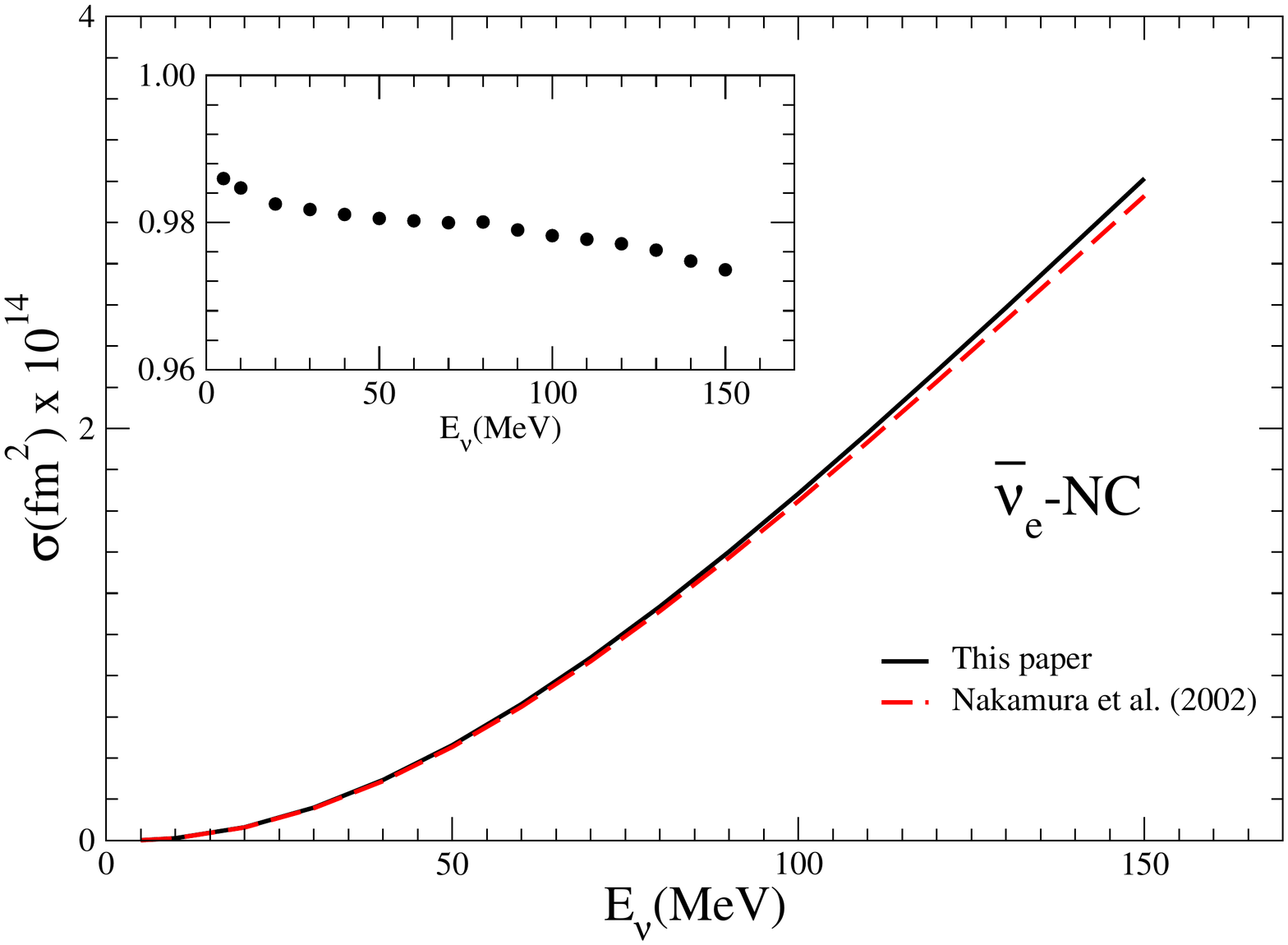}}
\caption{Same as in Fig.~\ref{fig:nu-pp} but for the $\overline{\nu}_e$-NC induced process on the deuteron.}
\label{fig:nub-np}
\end{figure}

\begin{figure}[bth]
\centerline{\includegraphics[width=16cm]{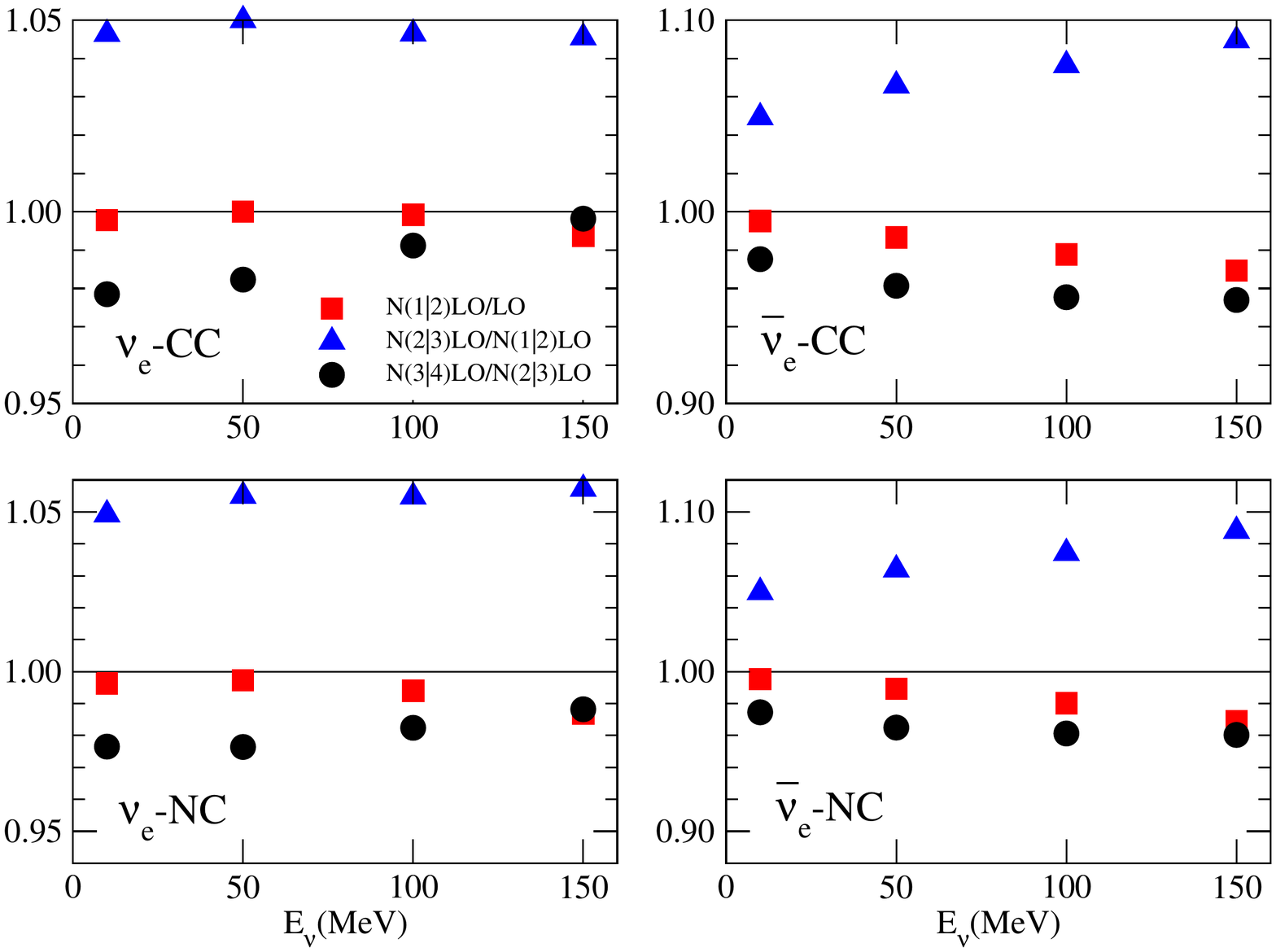}}
\caption{(Color online) The convergence pattern as function of increasing order in the chiral
expansion of the weak current. Ratios of corrections at a given order relative to the preceding order are shown.
Note that the $y$-axis scale in the  r.h.s.~panels is doubled relative to that in the l.h.s.~panels. }
\label{fig:ratio}
\end{figure}

Total cross sections, integrated over the final lepton energy and scattering angle
and obtained for the $\nu_e$-CC, $\overline{\nu}_e$-CC,  $\nu_l$-NC, and
$\overline{\nu}_l$-NC processes, are shown, respectively, in Figs.~\ref{fig:nu-pp}--\ref{fig:nub-np},
where they are compared to the corresponding predictions from Ref.~\cite{Nakamura02}
for incoming neutrino energies ranging from threshold up to 150 MeV.  The present $\chi$EFT
calculations are based on the Entem and Machleidt potentials of Refs.~\cite{Entem2003,Machleidt2011}
corresponding to cutoffs $\Lambda\,$=$\, 500$ and 600 MeV, 
and weak (vector and axial) current and charge operators of Refs.~\cite{Pastore2009,Pastore2011,Piarulli2013,Baroni2016},
as described in the previous section.
Matrix elements of these operators, suitably regularized as in Sec.~\ref{sec:lecs},
between the initial deuteron and final two-nucleon scattering states are evaluated with
the methods developed in Ref.~\cite{Shen2012}.  In practice, this entails obtaining the
two-nucleon radial wave functions from solutions of the Lippmann-Schwinger equation
in pair spin-isospin $ST$ channels with total angular momentum $J \le J_{\rm max}$, and
in approximating these radial wave functions by spherical Bessel functions in channels
with $J > J_{\rm max}$.  The full wave function, labeled by the relative momentum
${\bf p}$ (and corresponding energy $p^2/(2\mu)$, $\mu$ being the reduced mass) and
discrete quantum numbers $ST$, is then reconstructed from its partial-wave expansion~\cite{Shen2012}.
Consequently, interaction (including Coulomb in the case of two protons) effects in the final
scattering states are exactly accounted for only in channels with $J\le J_{\rm max}$.
For the neutrino energies of interest here, however, we find that these effects are negligible when
$J_{\rm max} \gtrsim 5$~\cite{Shen2012}.
The cross sections increase rapidly, by over two orders of
magnitude, as the neutrino energy increases from threshold
to 150 MeV.  Nevertheless, the present $\chi$EFT predictions
remain close to, albeit consistently larger at the 1--2\% level than,
those obtained in the conventional frameworks of Refs.~\cite{Nakamura02} and~\cite{Shen2012},
as shown explicitly for the case of Ref.~\cite{Nakamura02} by the insets in Figs.~\ref{fig:nu-pp}--\ref{fig:nub-np}.
The present $\chi$EFT electro-weak current and the meson-exchange models adopted
in Refs.~\cite{Nakamura02} and~\cite{Shen2012} provide an excellent description of
low-energy observables in the two- and three-nucleon systems (see Refs.~\cite{Piarulli2013,Marcucci2016}
and references therein).  In particular,
the axial current in both approaches ($\chi$EFT and meson-exchange) is
constrained to reproduce the tritium Gamow-Teller matrix element.
\begin{table}[bth]
\caption{Total cross sections in fm$^2$, corresponding to cutoff $\Lambda=\,$500 MeV,
for the CC-induced processes on the deuteron at selected initial neutrino energies and
at increasing orders in the chiral counting.  Referring to Figs.~\ref{fig:f2}--\ref{fig:f5a}, the
rows are labeled as follows: LO for the leading-order vector and axial current and charge;
N($1|2$)LO including the vector current and axial charge at N1LO, and the axial current
and vector charge at N2LO; N($2|3$)LO including the vector current at N2LO, and the
axial current and vector charge at N3LO; N($3|4$)LO including the vector current and
axial charge at N3LO, and the axial current and vector charge at N4LO.  Also listed are
the results at LO and N($3|4$)LO but $\Lambda=600$ MeV (labeled as LO$^\star$ and N($3|4$)LO$^\star$),
and those obtained in the conventional frameworks of (i) Ref.~\cite{Shen2012}
in impulse approximation (IA) and with inclusion of two-body currents (TOT)
and (ii) Ref.~\cite{Nakamura02} with inclusion of two-body currents (TOT).  The notation
$(xx)$ means $10^{xx}$.} 
\begin{tabular}{c||c|c|c|c||c|c|c|c||}
          & \multicolumn{4}{c} {$\sigma(\nu_e$-CC)} & \multicolumn{4}{c} {$\sigma(\overline{\nu}_e$-CC)}  \\
\hline\hline
 $E_\nu$ (MeV)  & 10 &  50  & 100  & 150 & 10 & 50 &  100 & 150 \\
    \hline
     LO            & 2.676(--16) & 1.345(--14)   &  6.611(--14)    & 1.591(--13)     & 1.243(--16)      &  7.441(--15)     & 2.661(--14)        &   4.944(--14)     \\
 N($1|2$)LO       &2.670(--16)   & 1.345(--14)  & 6.606(--14)   & 1.581(--13)    &  1.237(--16)     &  7.341(--15)     &  2.602(--14)        &  4.792(--14)      \\\
 N($2|3$)LO       &   2.794(--16) & 1.413(--14)     &6.913(--14)      & 1.653(--13)    &   1.298(--16)    &7.825(--15)         &  2.801(--14)        &5.221(--14)        \\
 N($3|4$)LO       & 2.734(--16)   &  1.388(--14)     &   6.852(--14)   &  1.650(--13)    &   1.266(--16)    &   7.523(--15)    &  2.676(--14)        & 4.981(--14)      \\
 \hline
LO$^\star$ &2.666(--16)    & 1.342(--14)   &6.593(--14)   &  1.588(--13)  &  1.239(--16)   & 7.417(--15)    &  2.653(--14)   & 4.925(--14)    \\
 N($3|4$)LO$^\star$ &2.729(--16)    & 1.388(--14)   &6.858(--14)   &  1.656(--13)  &  1.263(--16)   & 7.520(--15)    &  2.679(--14)   &4.998(--14)    \\
 \hline
 IA Ref.~\cite{Shen2012} & 2.630(--16)   &  1.314(--14)      & 6.424(--14)     & 1.516(--13)    & 1.219(--16)       & 7.260(--15)        & 2.567(--14)      &  4.688(--14)     \\
 TOT Ref.~\cite{Shen2012} & 2.680(--16)   &  1.348(--14)      &  6.631(--14)     &  1.574(--13)    & 1.242(--16)       &7.403(--15)        &2.606(--14)      & 4.751(--14)     \\
 \hline
 TOT Ref.~\cite{Nakamura02} & 2.708(--16)   &1.376(--14)      &  6.836(--14)     &  1.641(--13)    &  1.242(--16)       & 7.372(--15)        & 2.618(--14)      &  4.871(--14)     \\
 \hline\hline
\end{tabular}
\label{tab:cppp}
\end{table}
The $\chi$EFT cross sections of Figs.~\ref{fig:nu-pp}--\ref{fig:nub-np} correspond
to cutoff $\Lambda\,$=$\, 500$, but their variation as $\Lambda$ is increased
to 600 MeV remains well below 1\% over the whole energy range, as can be seen
in Tables~\ref{tab:cppp} and~\ref{tab:cpnp}, rows labeled N($3|4$)LO and N($3|4$)LO$^\star$.
The convergence of the chiral expansion is also shown in these tables, where
the various rows are labeled in accordance with the power counting adopted in the
present work, see Figs.~\ref{fig:nu-pp}--\ref{fig:nub-np}.  A graphical representation
of this convergence is provided by Fig.~\ref{fig:ratio}.  Overall, contributions beyond LO
lead to a couple of \% increase in the cross sections for both the CC and NC processes.
A similar increase due to two-body terms in the weak current is obtained
in the conventional calculations, see rows labeled IA and TOT in
Tables~~\ref{tab:cppp} and~\ref{tab:cpnp}.  Note that the IA row corresponds to results obtained with
one-body currents, including relativistic corrections~\cite{Shen2012}.   These IA currents
are the same as the $\chi$EFT ones illustrated by panel (a) of Fig.~\ref{fig:f2},
panels (a) and (b) of Fig.~\ref{fig:f5}, panels (a)-(d) of Fig.~\ref{fig:f2a}, and
panel (a) of Fig.~\ref{fig:f5a}.  Since the contributions due to the OPE
two-body terms in the vector current, panels (b) and (c) of Fig.~\ref{fig:f2},
and axial charge, panels (b) and (c) of Fig.~\ref{fig:f5a}, are very small,
then the difference between the IA and N($1|2$)LO results essentially 
reflects differences in the wave functions obtained from conventional
and chiral potentials.  Indeed, the overall $\sim 2$ \% offset between
the TOT and N($3|4$)LO predictions is primarily due to these differences.
\begin{table}[bth]
\caption{Same as in Table~\ref{tab:cppp} but for the NC-induced processes.}
\begin{tabular}{c||c|c|c|c||c|c|c|c||}
          & \multicolumn{4}{c} {$\sigma(\nu_e$-NC)}& \multicolumn{4}{c} {$\sigma(\overline{\nu}_e$-NC)}  \\
\hline\hline
 $E_\nu$ (MeV)  & 10 &  50  & 100  & 150 & 10 & 50 &  100 & 150 \\
    \hline
    LO            &  1.101(--16) & 5.872(--15)   &2.660(--14)      &5.991(--14)      & 1.050(--16)      &  4.554(--15)       & 1.664(--14)        & 3.175(--14)       \\
 N($1|2$)LO       & 1.097(--16)  & 5.856(--15)     & 2.644(--14)     &  5.912(--14)    &    1.045(--16)   & 4.505(--15)        & 1.631(--14)        &  3.076(--14)      \\
 N($2|3$)LO       &  1.151(--16) & 6.178(--15)    &  2.789(--14)    &  6.250(--14)    &    1.097(--16)   &   4.793(--15)      &  1.752(--14)       &  3.347(--14)      \\
 N($3|4$)LO       & 1.124(--16)  &  6.032(--15)   &   2.740(--14)   &  6.176(--14)    &  1.069(--16)     &  4.625(--15)      &    1.684(--14)      &  3.214(--14)      \\
 \hline
LO$^\star$ & 1.096(--16)     &5.853(--15)    & 2.652(--14)  & 5.973(--14)    &  1.045(--16)  & 4.539(--15)   & 1.659(--14)     & 3.165(--14)   \\
 N($3|4$)LO$^\star$ & 1.121(--16)     & 6.028(--15)    & 2.742(--14)  &6.191(--14)    &   1.067(--16)  & 4.622(--15)   &1.685(--14)     &3.224(--14)   \\
 \hline
IA Ref.~\cite{Shen2012} & 1.084(--16)   & 5.747(--15)     &  2.577(--14)     &5.720(--14)   & 1.033(--16)       & 4.449(--15)        & 1.604(--14)      & 3.003(--14)     \\
TOT Ref.~\cite{Shen2012} & 1.104(--16)   &5.892(--15)        &  2.657(--14)     &  5.935(--14)     &   1.053(--16)     &4.546(--15)        &1.640(--14)        &    3.075(--14)  \\
\hline
TOT Ref.~\cite{Nakamura02} & 1.107(--16)   &  5.944(--15)     &  2.711(--14)     & 6.130(--14)  &  1.053(--16)       & 4.535(--15)        & 1.647(--14)      &3.129(--14)     \\
 \hline\hline
\end{tabular}
\label{tab:cpnp}
\end{table}

The cross sections for the $\nu_l$-NC and $\overline{\nu}_l$-NC processes
only differ in the sign of the interference response function $R_{xy}$ in Eq.~(\ref{eq:xswa}).
In the case $\nu_e$-CC and $\overline{\nu}_e$-CC processes, additional
differences result from isospin-symmetry breaking terms in the final state
interactions of $pp$ versus $nn$.  At low energies ($E_\nu \lesssim 10$ MeV),
cross sections are dominated by the axial current, the associated contributions
being more than two orders of magnitude larger than those from the
vector current.  As the energy increases, vector-current contributions increase
becoming comparable, albeit still significantly smaller by over a factor
of five at $E_\nu=150$ MeV than, axial-current ones.  As a consequence,
the $\nu_l$-NC and $\overline{\nu}_l$-NC are fairly close at low energies,
but diverge significantly from each other as the energy increases.  Because of the aforementioned
isospin-symmetry breaking effects (primarily induced by the Coulomb
repulsion), the $\nu_e$-CC and $\overline{\nu}_e$-CC differ even
at low energies.  Finally, contributions from the axial charge
are negligible at $E_\nu \sim 10$ MeV, since at those energies the cross section
is dominated by the $^1$S$_0$ channel, to which axial-charge transitions from the $^3$S$_1$-$^3$D$_1$
state of the deuteron are strongly suppressed.  However, these axial-charge contributions remain well below 1\%
even at the high end of the energy range studied in this work, $E_\nu=150$ MeV.
At this latter energy, for example, ignoring these axial-charge contributions altogether would
reduce the $\nu_l$-NC ($\overline{\nu}_l$-NC) cross section from the N($3|4$)LO value of
6.176 (3.214) listed in Table~\ref{tab:cpnp} to 6.157 (3.194) in units of $10^{-14}$ fm$^2$. 
Thus, uncertainties in the values of the LECs $c_{5,2}$ and $c_{5,3}$ in the contact axial
charge do not have a significant impact on the present cross section predictions.

\begin{figure}[bth]
\centerline{\includegraphics[width=16cm]{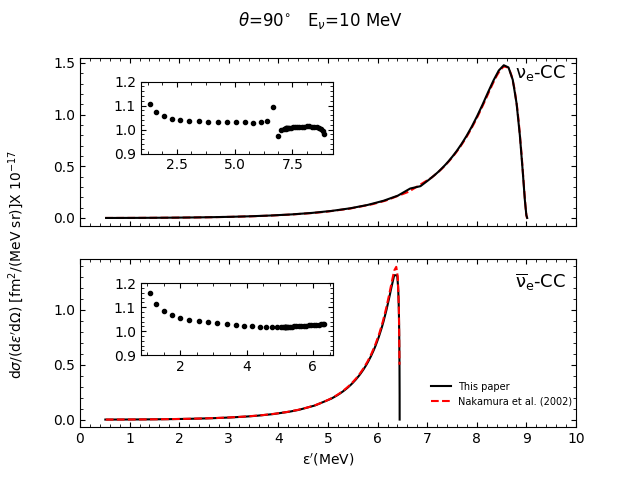}}
\caption{  (Color online) Double differential cross sections in ${\rm fm}^2/({\rm MeV}\,{\rm sr})$ for the $\nu_e$-CC and $\overline{\nu}_e$-CC induced process on the deuteron.
The solid line corresponds to the $\chi$EFT calculation with cutoff $\Lambda=$ 500 MeV, based on the chiral
potential of Ref.~\cite{Entem2003} and including electro-weak contributions up to N3LO in the vector current
and axial charge, and up to N4LO in the axial current and vector charge, see Figs.~\ref{fig:f2}--\ref{fig:f5a}.
The dashed line is obtained within the conventional meson-exchange picture of Ref.~\cite{Nakamura02}.
The inset shows the ratio of conventional to $\chi$EFT predictions.}
\label{fig:diff_sigma}
\end{figure}

Finally, for the purpose of illustration, Fig.~\ref{fig:diff_sigma} shows the double-differential
cross sections for CC-$\nu_e$ and CC-$\overline{\nu}_e$ induced processes as function of the final lepton
energy at a fixed scattering angle of $90^{\circ}$ and for incident neutrino energy of $10$ MeV.  The deuteron
wave functions are obtained from the N3LO chiral potential with cutoff $\Lambda=500$ MeV.  The energy
spectrum of the $\chi$EFT predictions closely matches that of Ref.~\cite{Nakamura02}.  We have not explicitly
verified that this agreement persists for different combinations of final lepton scattering angles and incident
neutrino energies.  However, we expect this to be the case for both the CC and NC reactions.

\section{Summary and Conclusions}
\label{sec:concl}
Cross sections for the reactions $^2$H($\nu_e,e^-$)$pp$, $^2$H($\overline{\nu}_e,e^+$)$nn$,
and $^2$H($\nu_l/\overline{\nu}_l,\nu^{\,\prime}_l/\overline{\nu}^{\,\prime}_l$)$np$ have
been studied in $\chi$EFT with the chiral potentials of Refs.~\cite{Entem2003,Machleidt2011}
and chiral electro-weak current of Refs.~\cite{Pastore2009,Pastore2011,Piarulli2013,Baroni2016,Baroni2016a}.
The potentials include intermediate- and long-range parts mediated by one- and multi-pion
exchanges, and a short-range part parametrized in terms of contact interactions, whose LECs
have been constrained by fits to the nucleon-nucleon database for energies ranging from zero
up to the pion-production threshold.  The vector- and axial-vector components of the weak
current have been derived up to one loop and include primarily one- and two-pion exchanges.  In addition
to these loop corrections, a number of contact terms occur.  The five LECs that multiply the
contact currents in the vector sector have been determined by a combination of resonance-saturation
arguments and fits to photo-nuclear data in the two- and three-nucleon systems.  Five LECs also enter
the axial sector (in the limit of low-momentum transfer processes).  Four of these
are in the charge operator: two are known from analyses of $\pi N$ data, while
the remaining two have yet to be determined and, in the present work, have been assumed
to be of natural size.  However, it is worthwhile emphasizing that the neutrino cross sections under
consideration are only marginally impacted by the axial-charge components in the weak
current.  The fifth and only LEC entering the axial current has been fixed by reproducing the tritium
Gamow-Teller matrix element.

Higher order contributions beyond LO lead to an overall increase by about 2\% in the
cross sections obtained with LO transition operators.  Predictions are also fairly insensitive
to variations in the short-range cutoff $\Lambda$, and change by a few parts in a thousand
as $\Lambda$ is changed from 500 to 600 MeV in both the potential and weak current.  As
illustrated by Fig.~\ref{fig:ratio}, there is good convergence in the chiral expansion of the weak
current.  The $\chi$EFT cross-section predictions reported here are consistently larger by a
couple of percent than corresponding results obtained in conventional formulations based on
meson-exchange phenomenology~\cite{Nakamura02,Shen2012}.  These conventional
calculations too are based on a model for the electro-weak current that provides an excellent
description of electromagnetic observables in the few-nucleon systems and the tritium
$\beta$-decay rate; indeed, two-body meson-exchange terms in the axial current are
constrained to reproduce the Gamow-Teller matrix element, just as in $\chi$EFT.  The
enhancement in the cross section due to (two-body) meson-exchange terms in the weak
current is similar (about 2\%) to that obtained in the $\chi$EFT calculations.  Indeed,
as noted in the previous section, the approximately 1--2\% offset between the conventional
(Refs.~\cite{Nakamura02,Shen2012}) and present $\chi$EFT results originates from
differences in the deuteron and two-nucleon continuum wave functions obtained with the
corresponding potentials rather than from the modeling of the weak current.  To explore
this point further, we have carried out preliminary calculations of the $\nu_e$-NC and
$\overline{\nu}_e$-NC cross sections with the LO weak current using one of the recently
developed ``minimally non-local'' configuration-space chiral potential of Ref.~\cite{Piarulli2015}.
We find that the LO $\nu_e$-NC ($\overline{\nu}_e$-NC) cross sections, in units of fm$^2$,
are $1.101 (1.050) \times 10^{-16}$ at 10 MeV and $5.937(3.147) \times 10^{-14}$ at 150 MeV,
to be compared, respectively, to $1.101(1.050) \times 10^{-16}$  and $5.991(3.175)\times 10^{-14}$
obtained with the chiral (and strongly non-local in configuration space) potentials of
Refs.~\cite{Entem2003,Machleidt2011} adopted in the present work.  This suggests
that the cross-section predictions based on chiral potentials and currents may have a
very small error ($< 1\%$) in the low-energy regime. A more rigorous way to estimate the theoretical uncertainty of the calculated cross-sections is described in Refs.~\cite{Epelbaum16a, Devries17}.
 
Finally, we conclude by noting that radiative corrections for the CC and NC processes due to bremsstrahlung
and virtual photon and $Z$ exchanges have been evaluated by the authors of Refs.~\cite{Towner98,Kurylov02}
at the low energies ($\sim 10$ MeV) most relevant for the SNO experiment, which measured the neutrino
flux from the $^8$B decay in the sun.  In the case of the $^2$H($\nu_e,e^-$)$pp$, these corrections lead to an
enhancement of the tree-level cross sections calculated in the present work (and
in Refs.~\cite{Nakamura02,Shen2012}), which ranges from about 4\% in the threshold
region to about 3\% at the endpoint of the $^8$B $\nu_e$-spectrum---this enhancement is in fact
larger than that induced by contributions in the weak current of order higher than leading.  While
the present results are not expected to impact the $^8$B $\nu_e$-spectrum deduced by the SNO
measurements (as the inset of Fig.~\ref{fig:nu-pp}), they should nevertheless be helpful in reducing
the theoretical error in this inferred spectrum.
\clearpage
\acknowledgments
The work of R.S. is supported by the U.S. Department of Energy, Office
of Nuclear Science, under contract DE-AC05-06OR23177.
The calculations were made possible by grants of computing time
from the National Energy Research Supercomputer Center.
%%%%%%%%%%%%%%%%%%%%%%%%%%%%%%%

%
%
%
 
%
%
\end{document}